\documentclass[a4paper,11pt]{article}
\usepackage{pos}
\title{Prospects of detecting gamma-ray signal of dark matter interaction with the MACE telescope}
 \ShortTitle{Dark Matter studies with MACE}

\author*[a,b]{M. Khurana}
\author[a]{A. Pathania}
\author[a,b]{K. K. Singh}
\author[a]{C. Borwankar}
\author[c]{P. K. Netrakanti}
\author[a,b]{K. K. Yadav}

\affiliation[a]{Astrophysical Sciences Division, Bhabha Atomic Research Centre, Mumbai 400085, India}
\affiliation[b]{Homi Bhabha National Institute, Mumbai 400085, India}
\affiliation[c]{Nuclear Physics Division, Bhabha Atomic Research Centre, Mumbai 400085, India}

\emailAdd{mkhurana@barc.gov.in}
\emailAdd{apathania@barc.gov.in}
\emailAdd{kksastro@barc.gov.in}
\emailAdd{chinmay@barc.gov.in}
\emailAdd{pawankn@barc.gov.in}
\emailAdd{kkyadav@barc.gov.in}
\abstract{The MACE (Major Atmospheric Cherenkov Experiment) telescope has started its regular gamma-ray observations at Hanle in India. 
Located at an altitude of $\sim$ 4.3 km above sea level and equipped with a 21 m diameter large quasi-parabolic reflector, it has the 
capability to explore the gamma-ray sky in the energy range above 20 GeV with very high sensitivity. In this work, we present the results 
from the feasibility studies for searching high-energy gamma-ray signals from dark matter interaction in potential astrophysical environments. 
We study the impact of MACE response function and other instrumental characteristics to probe the velocity average interaction cross-section 
($<\sigma v>$) of the weakly interacting massive particles (WIMPs), expected from the thermal dark matter freeze-out during the decoupling era. 
We consider the presence of dark matter in the form of pure WIMPs in the mass range 200 GeV - 10 TeV to produce distinctive gamma-ray spectra 
through its self-annihilation into standard model particles using the Pythia simulation package. The convolution of gamma-ray spectra corresponding 
to different standard model channels with the MACE response function is used to estimate the upper limit on $<\sigma v>$ for 100 hours of expected 
MACE observation of Segue1 (a dwarf spheroidal galaxy)  which is a potential site of dark matter.}

\ConferenceLogo{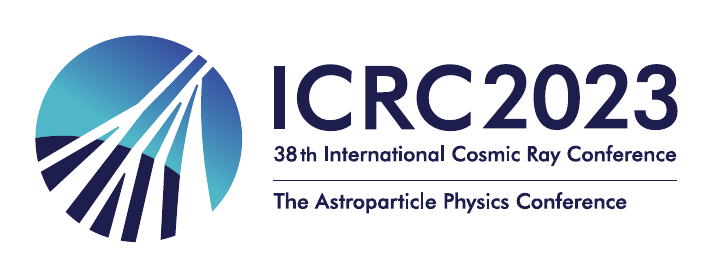}

\FullConference{%
38th International Cosmic Ray Conference (ICRC2023)\\
  26 July - 3 August, 2023\\
  Nagoya, Japan}
\begin{document}
\maketitle
%-----------------------------------------Section-1----------------------------------------------------------------------
\section{Introduction}\vspace{-0.2cm}
In modern cosmology, the concept of dark matter (DM) in Universe plays a dominant role to explain a wide range of astrophysical 
observations such as fluctuations in the cosmic microwave background radiation, baryon acoustic oscillations, flattening of galaxy 
rotation curves, gravitational lensing, large scale structure formations and their evolutions etc \cite{Bertone2018}. The ordinary or 
baryonic matter contributes $\sim$ 4$\%$ of the total energy budget of Universe while DM contributes $\sim$ 26$\%$ and remaining 70$\%$ 
is contributed by the dark energy. In spite of large DM abundance in galaxies and cluster of galaxies in the present Universe, its nature 
and particle origin, such as the mass and coupling to the standard model (SM) particles, remain the biggest mysteries of modern 
astro-particle physics and cosmology. In the widely accepted standard model of cosmology, a potential DM candidate is assumed to be 
neutral, long-lived, cold matter \cite{Bertone2005}. After decades of theoretical and experimental research on DM, it is generally 
believed that any theoretical description of DM should include a production mechanism which predicts its relic abundance 
of $\Omega_{DM} \approx 0.25$. Many theories, based on super-symmetric extensions of the SM of particle physics, predict weakly 
interacting massive particles (WIMPs) as natural potential candidates for cold dark matter in the mass range 10 GeV -10 TeV. 
An appealing frame-work for the genesis of WIMPs is thermal freeze-out from the primordial plasma of SM particles in the 
early Universe \cite{Lee1997}. Initially, WIMPs were in chemical and thermal equilibrium with the hot soup of SM particles produced after 
the big-bang and subsequently freezed-out of the thermal equilibrium as the interaction rate became smaller than the expansion rate of 
Universe. After this  freeze-out, the density of WIMPs remains constant. The expected value of velocity weighted average annihilation 
cross section ($<\sigma v>$) of WIMPs with mass of the electro-weak scale is $\approx~3 \times 10^{-26} cm^3 s^{-1}$ \cite{Roszkowski2018}.
Experimental searches of WIMPs approach the sensitivity to probe annihilation cross-section of this order to match the observed energy density 
of the cold dark matter.
\par
Direct and collider experiments for WIMP detection have not yielded any acceptable statistically significant signal so far. In recent years, 
indirect methods, related to the detection of signal from products of WIMP annihilation in the plausible astrophysical sites, have seen 
increasing interest. The products of WIMP annihilation can come from the hadronization or decay of primary SM particles such as quark-antiquark, 
boson and lepton through a given channel with its own branching ratio \cite{Cirelli2011}. The main type of products are neutrinos, positrons, 
anti-protons and gamma-rays. In this contribution, we present the feasibility study of search for gamma-ray signal of WIMP annihilation from 
the dwarf spheroidal galaxy Segue1 using the newly commissioned Major Atmospheric Cherenkov Experiment (MACE) telescope in India. 
This galaxy, located at a distance of $\sim$ 23 kpc from Earth, is considered to be one of the prominent DM sites with a mass-to-luminosity 
(M/L) ratio of $\approx$ 3400 $(M_\odot/L_\odot)$, where $M_\odot$ and $L_\odot$ are the mass and luminosity of Sun respectively. 
%-----------------------------------------------Section-2------------------------------------------------------------
\section{MACE Telescope}
\begin{figure}[h!]
  \begin{minipage}[b]{0.5\linewidth}
    \centering
    \includegraphics[scale=0.16]{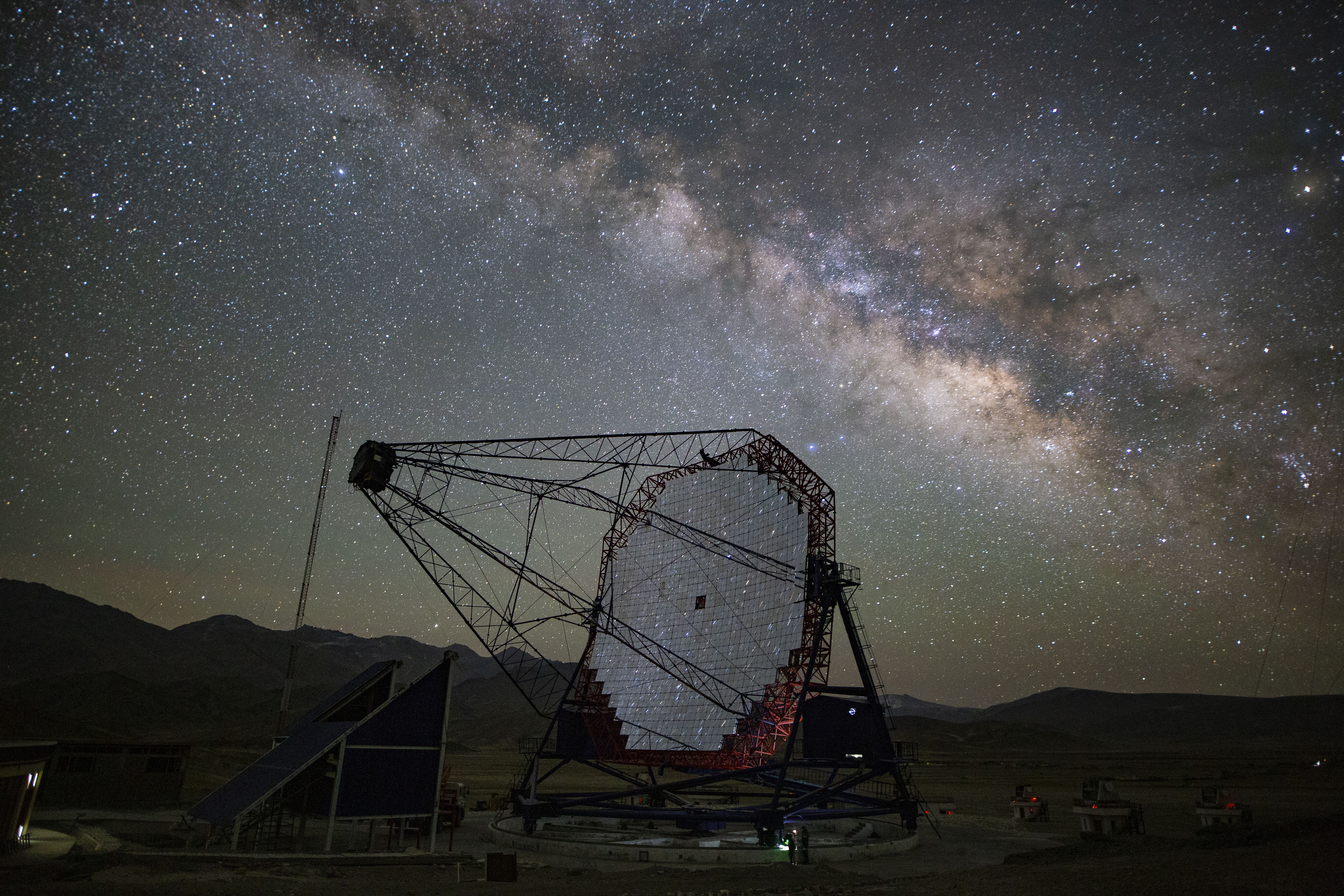}
    \captionof{figure}{The MACE Telescope at Hanle, Ladakh in India.}
     \label{fig:mace}	  
  \end{minipage}\hfill
  \begin{minipage}[b]{.5\linewidth}
    \centering
	  \begin{tabular}{||c|c||}
    
    \hline
     Altitude & 4.3 km (asl)  \\ [0.5ex] 
    \hline 
    Location & Hanle, India \\ 
     &  (32.8 $^\circ$ N, 78.9 $^\circ$ E)  \\ 
    \hline
    Diameter & 21m \\
    \hline
    Field of View & 4.36 $^\circ \times 4.03 ^\circ$ \\
    \hline
    Energy Range & Above 20 GeV \\
    \hline
    Energy Resolution & 20 $\%$ at 1 TeV \\  
    \hline
    Pixel Resolution & 0.125 $^\circ$ \\
    \hline
    Sensitivity & 2.7 $\%$ Crab Flux  \\
    & in 50 hrs \\
    \hline
\end{tabular}
    \captionof{table}{Salient features of the MACE telescope.}
    \label{tab-feature}
  \end{minipage}
\end{figure}

MACE is a newly commissioned imaging atmospheric Cherenkov telescope (IACT) at Hanle, Ladakh in India \cite{Yadav2022}.  
A picture of the telescope captured during night is shown in Figure \ref{fig:mace}. Located at an altitude 
of $\sim$ 4.3 km above sea level, MACE is the highest existing IACT in the world. The geographical location of the MACE telescope 
appropriately fills the longitudinal gap among the leading current and future IACTs around the globe \cite{Singh2021,Singh2022}. 
Equipped with a 21 m diameter large light collector, it has capability of detecting very high energy gamma-ray photons in the energy 
range above 20 GeV with very high point source sensitivity. The compact imaging camera at the focal plane comprises 1088-photomultipliers 
with a pixel resolution of 0.125$^\circ$ and provides a total field of view of $\sim~4.36^\circ~\times~4.03^\circ$. The important salient 
features of MACE are summarized in Table \ref{tab-feature}. These features place the MACE telescope as the second largest stand alone 
IACT after LST-1 in the northern hemisphere. It is a prime case for the installation of low threshold energy stereoscopic IACT systems 
at high altitude in future. 
%----------------------------------------Figure 1-MACE----------------------------------------- 
%---------------------------------------Section-3----------------------------------------
\section{$\gamma -$ ray signal from WIMP annihilation}
WIMPs in dense astrophysical environments would annihilate into a pair of SM particles and produce gamma-rays in final 
state from their hadronization, radiation and decay \cite{Bringmann2012}. These high energy photons can be eventually 
detected by the space- and ground-based telescopes ( e.g. MACE) provided the mass of WIMPs ($m_\chi$) is high enough to 
obtain the most robust and stringent constraints on WIMP annihilation into a variety of final states and therefore probing 
the parameter space of WIMPs as dark matter candidate. The gamma-ray continuum in astrophysical systems can be produced 
by the WIMP annihilation into any SM particle pair ($\chi + \chi \rightarrow SM + \bar {SM}$) with gamma-ray photons being 
either radiated off or emanating from the decay of a particle that has arisen either directly from a primary decay product 
or as a result of its hadronization. The photon energy ranges from $E_{max} \sim m_\chi$ to all the way down. 
The expected signal from the Majorana WIMP annihilation is characterized by a specific spatial shape and specific energy spectrum. 
The differential flux of gamma-ray photons lying within a solid angle $\Delta \Omega$ is expresses as \cite{Fermi2010}
\begin{equation}\label{WIMP_diff_eq}
    \frac{dN}{dE} = \frac{<\sigma v>}{8 \pi m_{\chi}^2} \sum_{f} b_{f} \frac{dN_{\gamma}^{f}}{dE} \times J(\Delta \Omega)   
\end{equation}
where $\frac{dN_{\gamma}^{f}}{dE}$ is the differential photon spectrum per WIMP annihilation event in the channel $`f'$ 
with its corresponding branching ratio $b_f$. The first term in the right hand side of Equation \ref{WIMP_diff_eq} accounts for the 
particle physics contributions and is therefore referred to as the \emph{particle physics factor}. It controls the shape of gamma-ray 
spectrum and carries the physics involved in WIMP interactions. The second term $J(\Delta \Omega)$ is known as the \emph{Astrophysical-J factor}. 
It is defined as 
\begin{equation}\label{Jeq}
    J(\Delta \Omega) = \iiint\limits_{l.o.s} \rho^2(r) dl d\Omega
\end{equation}
where $\rho(r)$ represents the density profile of WIMPs in a given astrophysical environment. J-factor is a measure of DM density square 
integrated over the line-of-sight (l.o.s) $l$ and solid angle $\Delta \Omega$. The density of DM is assumed to spherically symmetric and 
depends only the radial distance $r$ from the center. The value of J-factor, a measure of total DM content, plays an important role in source 
selection as it decides overall strength of the signal.

%--------------------------------------------------Section-4-----------------------------------------------
\section{Results and Discussion}
Dwarf spheroidal galaxies, less massive than Milky-Way, are considered as potential targets for unambiguous detection of dark matter 
signal since they have negligible astrophysical gamma-ray background. This is because neither known gamma-ray sources such as supernova 
remnants/pulsar wind nebula nor gases acting as targets for cosmic rays are observed in spheroidal galaxies. Being small in size at distant 
locations, these galaxies appear as point sources for ground-based gamma-ray telescopes like MACE and therefore the instrumental background 
is also negligible. Motivated by these characteristics, we have considered Segue1 as a test case for MACE in the present study.  

%--------------------------- Fig 2 --------------------------------------------
\begin{figure}[h!]
\centering
\includegraphics[width=0.6\textwidth]{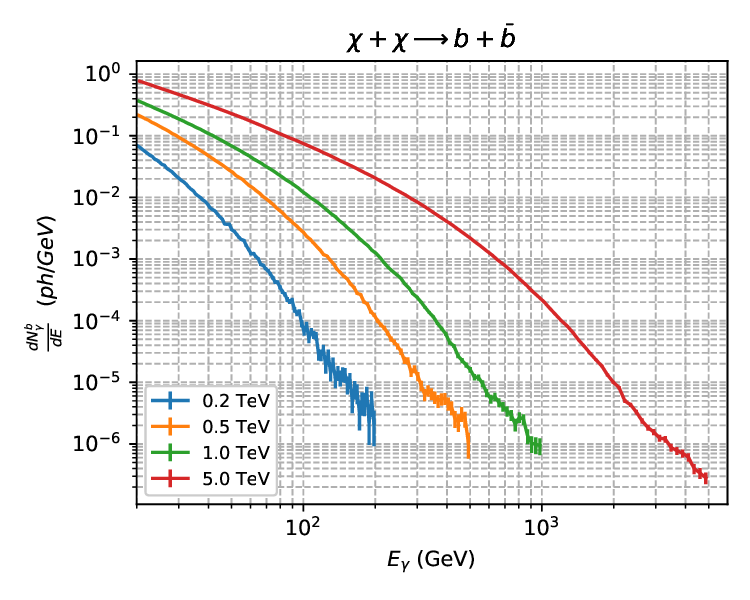}
\caption{Differential photon spectrum per WIMP annihilation in $b\bar{b}$-channel for different values of $m_\chi$.}
\label{fig:spectra}
\end{figure}

We have used Pythia numerical package \cite{Sjostrand2008} to estimate $\frac{dN_{\gamma}^{f}}{dE}$ by considering the Majorana nature of 
WIMPs and assuming that the WIMP pair annihilation  proceeds into a single channel alone with 100$\%$ probability. Based on the theoretical 
motivations available in literature, we have considered  $b\bar{b}, \tau^+ \tau^-, Z\bar{Z}, h\bar{h}$ \& $W^+ W^-$ as the single channel dominance 
for WIMP annihilation. A representative $\frac{dN_{\gamma}^{f}}{dE}$ for $b\bar{b}$ channel corresponding to different masses of WIMP is presented 
in Figure \ref{fig:spectra}. 

\begin{figure}[h!]
\centering
\includegraphics[width=0.6\textwidth]{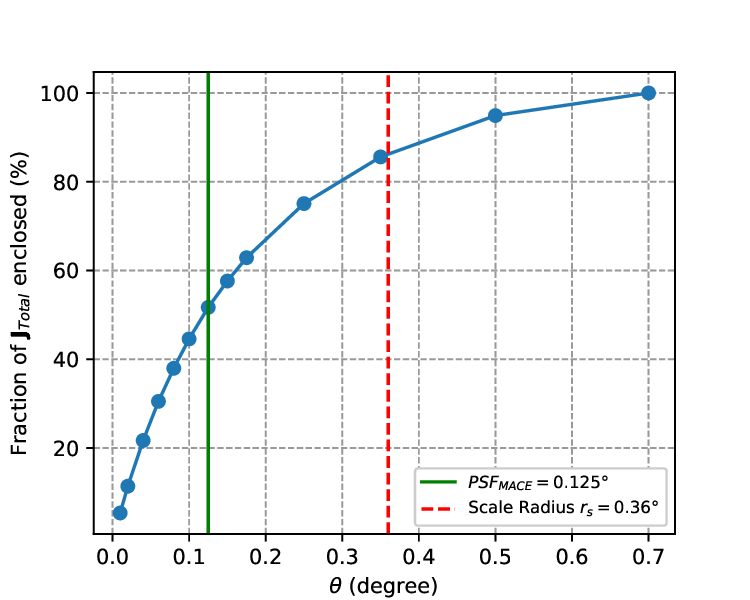}
\caption{Fraction of total J-factor enclosed as a function of angular extent ($\theta$) of the Segue1 galaxy.}
\label{fig:Jvstheta}
\end{figure}

Detailed physics involved in these interactions are either derived from theory or based on phenmological models with 
parameters derived from Large Hadron Collider data with strong nuclear forces governed by Quantum Chromodynamics. 
For estimation of J-factor, we consider the widely used dark matter density profile given by Navarro-Frenk-White (NFW) as \cite{Navarro1997}
\begin{equation}\label{NFW}
    \rho(r) = \frac{\rho_s}{\left( \frac{r}{r_s} \right) \left( 1 + \frac{r}{r_s} \right)^2}
\end{equation}
where $\rho_s$ and $r_s$  are defined as scale density and scale radius respectively. For Segue1, $\rho_s = 21.005 GeV/cm^3$ and $r_s = 0.1461 kpc$.  
Since the J-factor is calculated within an angular region of radius equal to the point spread function (PSF) of a given telescope, it is evident 
from Figure \ref{fig:Jvstheta} that majority of dark matter in Segue1 is contained within the MACE-PSF of $\approx 0.125^\circ$ for the scale 
radius of $r_s = 0.36^\circ$. Therefore, the value of J-factor for MACE is estimated to be $J_{MACE} = 3.05 \times 10^{19} GeV^2/cm^5$. 

%--------------------------- Fig 3 --------------------------------------------

\begin{figure}[h!]
\centering
\includegraphics[scale=0.60]{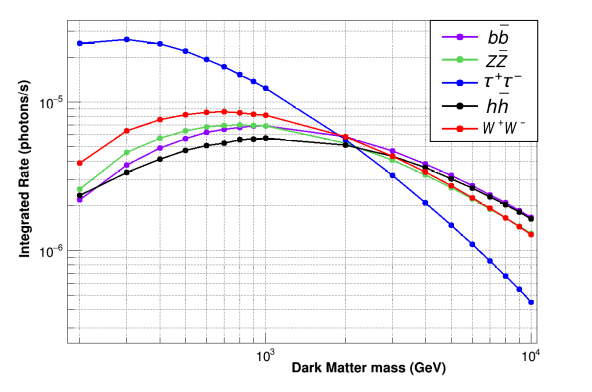}
\caption{Integral photon rate for different channels in MACE energy range as function of WIMP mass.}
\label{fig:IntRate}
\end{figure}

\begin{figure}[h!]
\centering	
\includegraphics[scale=0.60]{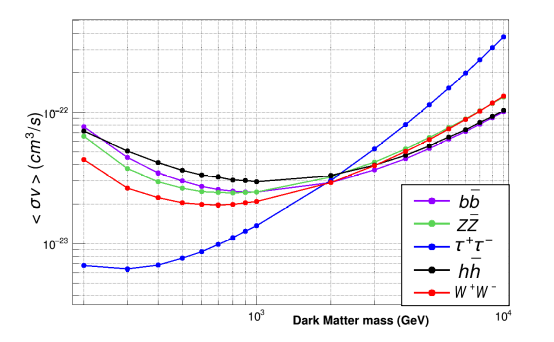}
\caption{Constraints on $<\sigma v>$ as a function of WIMP mass.}
\label{fig:sigma}
 \end{figure}

The integral gamma-ray rate from the Segue1 galaxy can be calculated by convoluting Equation \ref{WIMP_diff_eq} with the MACE response function $A_{eff}(E)$ and is given by 
\begin{equation}\label{eqn-rate}
	R_\gamma~=~ \int_{20 GeV} \frac{dN}{dE} A_{eff} (E) dE 
\end{equation}
The estimated integral gamma-ray photon rates as a function of WIMP mass ($m_\chi$) in the MACE energy range above 20 GeV for different channels of 
WIMP annihilation for $< \sigma v > \approx~3 \times 10^{-26} cm^3 s^{-1}$ are reported in Figure \ref{fig:IntRate}.
%------------------------------Figure-4-------------------------------------------
However, the value of expected integral gamma-ray rate ($R_{\gamma}^*$) from any weak source for 3$\sigma$ detection in a given observation ($T_{obs}$) can be 
obtained using the Li-Ma formula \cite{Li1983} and is given as
\begin{equation}\label{eqn-ER}
	R_{\gamma}^* = \frac{9 + \sqrt{81 + 72 R_{CR}T}}{2T}
\end{equation}
where $R_{CR}$ is the measured cosmic ray background rate by any ground-based telescope. Therefore, this is translated into an upper limit on $<\sigma v>$ for WIMP 
annihilation into a given channel. From Equations \ref{WIMP_diff_eq} \& \ref{eqn-rate}, the derived upper limit on  $<\sigma v>$ at 3$\sigma$ confidence level 
can be expressed as 
\begin{equation}
	<\sigma v> = 3 \times 10^{-26} \times \left( \frac{R_{\gamma}^*}{R_\gamma}\right)~~~~ \rm cm^3~s^{-1} 
\end{equation}
Expected results for $T_{obs} =$ 100 Hours using MACE observations of Segue1 galaxy at 3$\sigma$ confidence level are depicted in Figure \ref{fig:sigma}.
We observe that the value of $<\sigma v>$ can be constrained to an upper limit of $\sim$ 6.44$\times10^{-24} ~cm^3~s^{-1}$ at 3$\sigma$ confidence level 
for the WIMP annihilation of masses $m_\chi =$ 300 GeV through $\tau^+ \tau^-$ channel. The corresponding upper limit on the relic density parameter of 
WIMPs can be estimated using the relation 
\begin{equation}
	\Omega_{WIMP} = \frac{6.0 \times 10^{-27} \rm cm^3~s^{-1}}{<\sigma v>}
\end{equation}	
The value of $\Omega_{WIMP} \approx$ 0.06 is referred to as \emph{WIMP miracle}. Therefore, in order to arrive the so called WIMP miracle, the $<\sigma v>$ value 
should be obtained of the order of $10^{-25} \rm cm^3~s^{-1}$ from any observational approach. Long-term observations of Segue1 for about 160 Hours with  
the stereoscopic MAGIC telescope have placed stringent upper limit of $1.4 \times 10^{-23} \rm cm^3~s^{-1}$ (95$\%$ confidence level) on $<\sigma v>$ at 
$m_\chi =$ 700 GeV under the frame-work of brane-world theory \cite{Miener2022}. In this model of dark matter, the characteristics of the proposed massive 
brane fluctuations or branons match with those of the WIMPs \cite{Cembranos2003}.
%-----------------------------------------------------Figure 5----------------------------------------------------

%-------------------------------------------Section-5--------------------------------------------
\section{Conclusions}
Ground-based IACTs with extraordinary point source sensitivity are deemed to be an ideal instrument for searching weak gamma-ray signals 
produced from the WIMP annihilation in the dwarf spheroidal galaxies. In this study, we find that the recently commissioned MACE telescope 
can be effectively used to constrain the parameter space of WIMPs in the mass range 200 GeV - 10 TeV. The feasibility study suggest that 
100 Hours of dedicated observation of the Segue1 galaxy with MACE would be able to constrain $<\sigma v>$ value close to the order 
of $10^{-24} cm^3 s^{-1}$ at 3$\sigma$ confidence level for WIMP annihilation of mass $\sim$ 300 GeV into $\tau^+ \tau^-$. However, results 
corresponding to other decay channels such as $b\bar{b}, Z\bar{Z}, h\bar{h}$ \& $W^+ W^-$ can constrain the $<\sigma v>$ value up to the order 
of $10^{-23} cm^3 s^{-1}$.  

%---------------------------------------References--------------------------------------------------------


\begin{thebibliography}{99}
\bibitem{Bertone2018}
	G. Bertone, \emph{History of dark matter}, \emph{REVIEWS OF MODERN PHYSICS} {\bf 90} (2018) 045002
\bibitem{Bertone2005}
	G. Bertone et al., \emph{Particle dark matter: evidence, candidates and constraints}, \emph{Physics Reports} {\bf 405} (2005) 279
\bibitem{Lee1997}
	B. W. Lee \& S. Weinberg, \emph{Cosmological Lower Bound on Heavy Neutrino Masses}, \emph{Physical Review Letters} {\bf 39} (1997) 165
\bibitem{Roszkowski2018}
	L. Roszkowski et al., \emph{WIMP dark matter candidates and searches{\textemdash} current status and future prospects}, 
		\emph{Reports on Progress in Physics} {\bf 81} (2018) 066201
\bibitem{Cirelli2011}
	M. Cirelli et al., \emph{PPPC 4 DM ID: a poor particle physicist cookbook for dark matter indirect detection}, \emph{Journal of Cosmology 
		and Astroparticle Physics} {\bf 2011} (2011) 051
\bibitem{Yadav2022}
	K. K. Yadav et al., \emph{Commissioning of the MACE gamma-ray telescope at Hanle, Ladakh, India}, \emph{Current Science} {\bf 123} (2022) 
		1428-1435
\bibitem{Singh2021}
        K. K. Singh \& K. K. Yadav, \emph{20 Years of Indian Gamma Ray Astronomy Using Imaging Cherenkov Telescopes and Road Ahead}, 
	\emph{Universe} {\bf 7} (2021) 96
\bibitem{Singh2022}
	K. K. Singh, \emph{Gamma-ray astronomy with the imaging atmospheric Cherenkov telescopes in India}, \emph{Journal of Astrophysics \& Astronomy} 
	{\bf 43} (2022) 3
\bibitem{Bringmann2012}		
	T. Bringmann \& C. Weniger, \emph{Gamma Ray Signals from Dark Matter: Concepts, Status and Prospects}, \emph{Physics of Dark Universe} 
		{\bf 1} (2012) 194
\bibitem{Fermi2010}
	Fermi-LAT collaboration, \emph{Observations of Milky Way Dwarf Spheroidal galaxies with the Fermi-LAT detector and constraints 
		on Dark Matter models}, \emph{Astrophysical Journal} {\bf 712} (2010) 147
\bibitem{Sjostrand2008}		
	T. Sjostrand et al., \emph{A Brief Introduction to PYTHIA 8.1}, \emph{Computational Physics Communication} {\bf 178}  (2008) 852		
\bibitem{Navarro1997}
	J. F. Navarro et al., \emph{A Universal density profile from hierarchical clustering}, \emph{Astrophysical Journal} {\bf 490} (1997) 493
\bibitem{Li1983}
	T. P. Li \&  Y. Q. Ma, \emph{Analysis methods for results in gamma-ray astronomy}, \emph{Astrophysical Journal} {\bf 272} (1983) 317
\bibitem{Miener2022}
	T. Miener et al., \emph{Constraining branon dark matter from observations of the Segue 1 dwarf spheroidal galaxy with the MAGIC telescopes}, 
		\emph{Journal of Cosmology and Astroparticle Physics}, {\bf 05} (2022) 005
\bibitem{Cembranos2003}
	J. A. Cembranos et al., \emph{Brane-World Dark Matter}, \emph{Physical Review Letters} {\bf 90} (2003) 241301
\end{thebibliography}
\end{document}